# Mapping AI Ethics Narratives: Evidence from Twitter Discourse Between 2015 and 2022

Mengyi Wei[1], Puzhen Zhang[1], Chuan Chen [1]*, Dongsheng Chen[1], Chenyu Zuo[2], Liqiu Meng[1]

[1] *Chair of Cartography and Visual Analytics, Technical University of Munich*

[2] *Center for Sustainable Future Mobility (CSFM), ETH Zurich*

## Abstract

Public participation is indispensable for an insightful understanding of the ethics issues raised by AI technologies. Twitter is selected in this paper to serve as an online public sphere for exploring discourse on AI ethics, facilitating broad and equitable public engagement in the development of AI technology. A research framework is proposed to demonstrate how to transform AI ethics-related discourse on Twitter into coherent and readable narratives. It consists of two parts: 1) combining neural networks with large language models to construct a topic hierarchy that contains popular topics of public concern without ignoring small but important voices, thus allowing a fine-grained exploration of meaningful information. 2) transforming fragmented and difficult-to-understand social media information into coherent and easy-to-read stories through narrative visualization, providing a new perspective for understanding the information in Twitter data. This paper aims to advocate for policy makers to enhance public oversight of AI technologies so as to promote their fair and sustainable development.

Keywords: AI Ethics, Twitter, Narrative Visualization, Public Participation, Event Evolution Analysis, Social Media

## 1. Introduction

The essence of AI (Artificial Intelligence) technology is to serve human beings. However, the current development of AI is in the hands of technical and economic elites and lacks active public engagement (Corbett et al., 2023; Dollbo, 2023; Gilman, 2023). The public is reduced to consumers simply waiting for a new generation of AI technology. Therefore, when it comes to the ethical problems that AI technology poses for people, if public discourse and sentiment are ignored, the decision-making power on how to better deploy AI will be fall into special interests (Rainie, 2018). Often public discourse is ignored when using data. In that case, it is highly susceptible to becoming a means for companies to pursue profit, thereby infringing upon the public's privacy rights. In recent decades, research on AI ethics has evolved from a policymaker-based perspective (Jobin et al., 2019; Smuha, 2019) to one that increasingly values the contribution of AI experts (Pflanzer et al., 2023; Sanderson et al., 2023), the research direction is gradually turning to the general public. For example, studies now encompass interviews to gauge the perspectives of students and educators regarding ChatGPT usage (Iskender, 2023; Zhou et al., 2024). Public perceptions of various facets of AI were scrutinized using nationally representative survey data (N=2700), providing insight into individuals' views on the risks and benefits linked to AI (Bao et al., 2022). An online experiment was conducted to explore the relationship between ambiguous public perceptions of AI and people's willingness to use AI technology (Schwesig et al., 2023). However, research samples based on questionnaires and interviews are currently limited, making it difficult to significantly increase public participation. Therefore, it is necessary to listen to diverse public voices to ensure AI's fair and equitable development.

With the advent of the intelligent era, the content and information dimensions covered by social media data are becoming increasingly affluent, which offers the possibility of studying AI ethics from a public perspective to some extent (Manovich, 2012). As an online social platform, Twitter provides a convenient channel for the public to have an equal voice in the AI ethical discourse. However, little research provides a holistic perspective to analyze public discourse on AI ethics using Twitter data. On the one hand, social media data is often filled with a large amount of noise, making it challenging to extract meaningful information (Feng et al., 2023). Moreover, the information contained in social media data is often fragmented, making it difficult to understand. A readable and coherent narrative that can explicitly present the AI ethics discourse embedded in social media data is still missing.

Our work is dedicated to proposing a new framework for mining effective public voices on AI ethics discourse on social media. Two main parts are involved: First, we propose a neural network combined with a large language model to hierarchically structure AI ethics discourse on Twitter. By revealing trending topics of public discussion while taking into account small but important subtopics, the approach allows for a fine-grained exploration of meaningful information. Second, we realize the problem of people not seeing the world in front of them until they are in narrative mode (Sadler, 2018; E Segel and Heer, 2010). As information obtained from social media data is often fragmented and difficult to understood, we use narrative visualization to integrate scattered information and construct readable and coherent narratives through story maps and event evolution diagrams. In this way, we attempt to promote better understanding of AI ethics discourse on Twitter. In short, the following two research questions are addressed:

RQ1: How to extract effective public voices from large amounts of Twitter data related to AI ethics?

RQ2: How to transform fragmented Twitter information into a coherent, readable narrative?

The second section provides an overview of the relevant research, followed by an introduction to the research methods in the third section. The results of the two research questions are presented in the fourth section. The fifth section contains a discussion of the results. The limitations and implications of this study are discussed in the sixth section, followed by a conclusion in the final section.

## 2. Related Work

The literature review of this paper is structured into three aspects. Firstly, it delves into examining research about AI ethics within the context of Twitter data. Secondly, it outlines the methodologies employed for topic extraction in tweets, analyzing their merits and drawbacks. Lastly, it provides a comprehensive overview of visualization studies concerning social media data.

## 2.1 AI ethics in Twitter Data

While Twitter may not fully represent the entirety of the social media landscape (Dijck, 2013), it has emerged as a prominent platform for individuals to express opinions and engage in online discourse (Anger and Kittl, 2011). Additionally, it serves as a valuable research tool for scholars investigating various scientific topics (Chen et al., 2023; Fu et al., 2022; Hua et al., 2022). Research on AI ethics utilizing Twitter data has been focused either on specific topics in a domain, such as security, equity, and emotional sentiments in the blockchain domain, with the aim to address fairness concerns in blockchain design related to transaction ordering (Fu et al., 2022), or on a single category of ethical issues, such as using Twitter data to examine the relationship between individual privacy settings and self-disclosure on Twitter, considering cultural values across different contexts (Liang et al., 2017). Recent studies have also investigated public discussions and reactions to ChatGPT (Haque et al., 2022). Specifically, by analyzing tweets within a month of ChatGPT's launch, the study identified the most relevant topics and sentiments of the public, and ethical challenges to be addressed as ChatGPT develops (Taecharungroj, 2023). While such studies shed light on prevailing topics of public discourse and sentiment analysis on specific AI domains or ethical concerns, few provide a comprehensive analysis of AI ethics-related topics on Twitter.

Furthermore, research related to Twitter data often focuses on two types of analysis: trending public discussion topics and sentiment changes. For instance, in a blockchain study related to AI ethics, the top 30 keywords associated with #Flashbots and #MEV were extracted (Fu et al., 2022). The tweets about ChatGPT were categorized into nine topics, ranging from "Future Career & Opportunities" to "Disruptions for Software", and the corresponding sentiment distribution for each topic was presented in (Haque et al., 2022). Taecharungroj (2023) employed the LDA method to extract public topics on ChatGPT, mainly focusing on news, technology, and reactions, and identified five functional domains: creative writing, essay writing, prompt writing, code writing, and answering questions. Topic extraction and sentiment analysis played an important role in using Twitter data to reveal public viewpoints and expressions on online social platforms (Boon-Itt and Skunkan, 2020; Hua et al., 2022). However, this information represents only a small part of Twitter data and is often too fragmented to form coherent, readable information. This study focuses on extracting fine-grained and rich information from Twitter and attempts to present it in a more readable and coherent narrative, thereby providing a comprehensive perspective for the public to understand AI ethics.

## 2.2 Topic Mining

Topic extraction methods in Twitter data are mainly divided into network-based methods and text-based methods. Network-based topic classification methods include social network analysis, graph mining, and topic propagation models. Lee et al. (2011) identified the top five similar topics among 18 popular categories on Twitter based on the number of influential users in common and validated that network-based classification modeling methods can achieve up to 70% classification accuracy. Azam et al. (2015) proposed a social graph generation method, treating tweets as nodes and using the Markov clustering technique to decompose the social graph into various clusters, each corresponding to specific events, thereby achieving event classification. Huang and Mu (2014) employed a combination of clustering algorithms with hashtag propagation algorithms to detect topics on Twitter, to classify the tweets into different clusters and then using a label propagation mechanism to label tweets that overlap in different clusters. Finally, this method was compared with other clustering algorithms, validating the accuracy of the label-propagation-based algorithm.

Text-based topic extraction from Twitter data includes keyword extraction and topic modeling (Karami et al., 2020). A representative method for keyword extraction is the TF-IDF ( Term Frequency–Inverse Document Frequency). Alsaedi et al. (2016) proposed a novel temporal term TF-IDF method, which can overcome the drawbacks of traditional methods that require prior knowledge of the entire dataset by assuming that "words with higher frequencies in documents within specific periods are more likely to be selected for human-created document summaries," and validated the superiority of this method. With regard to topic modeling for the extraction of topics from Twitter data is often, LDA (Latent Dirichlet Allocation) is a commonly used method. Chen et al. (2023) used the LDA method to obtain topics related to "climate strikes" in Twitter data from 2018 to 2021, providing a reference for using social media to construct political issues and collective actions. However, these topic mining methods often suffer from a large amount of noise and a single structured nature, failing to effectively mine valuable information from social media data. Therefore, we proposed to combine neural networks with large language models, presenting hierarchical topic structure, and achieving fine-grained classification of social media data.

## 2.3 Narrative Visualization for Exploratory Data Analysis

Narrative visualization involves not only showcasing the data itself but embedding it in a broader context by constructing narrative content to reveal the stories and meanings behind the data (Chen et al., 2024; Correa and Silveira, 2022; E Segel and Heer, 2010). Sadler et al. (2018) pointed out that the challenge in Twitter-related communication research lies in how users understand the fragmented tweets presented to them and proposed creating coherent and meaningful mental narratives based on individual tweets combined with other materials to help readers create richer reading experiences. Their study addresses the issue of how fragmented tweets can provide more meaningful information, but text-based narratives alone may be insufficient to enhance readers' experiences in acquiring adequate information. Therefore, we use narrative visualization to present the tweet data information in this paper.

Narrative visualization, as a method for presenting textual information, has been widely applied in storytelling and social media to better convey information. Segel and Heer (2010) systematically reviewed the design space of narrative visualization in news media and identified seven types of narrative visualization genres, including magazine style, annotated chart, partitioned poster, flow chart, comic strip, slide show, and film/video/animation. Hullman et al. (2013) conducted a qualitative analysis of 42 professional narrative visualizations to explore the empirical knowledge of structure and sequence in narrative visualization. They proposed a graph-driven approach to identify effective sequences in visualizations and validated the functionality, their insight is useful for guiding narrative visualization and supporting visualization sequences. McKenna et al. (2017) investigated and identified flow factors in narrative visualization by analyzing 80 stories on websites and illustrating how they fit into the broader concept of narrative visualization. Concurrently, they conducted a crowdsourced study with 240 participants to explore the impact of different combinations of flow factors on reader engagement, ultimately finding that visual and navigational feedback can affect reader engagement, while control levels (e.g., discrete vs. continuous) do not. Metoyer et al. (2018) proposed an automatic method for generating text and visualization elements to study narrative visualization in online media, exploring users' reading experiences from the bidirectional perspective of text-to-visualization and visualization-to-text interaction.

Story mapping, as a form of narrative visualization, is a focal point of this study. A story map typically refers to a visual interpretation resembling the construction of semantic maps, webs, or networks, presenting geographically relevant information, events, or themes in a narrative format (Davis and McPherson, 1989; Freedman and Reynolds, 1980; Roth, 2021). Reutzel (1985) evaluated aspects such as reading comprehension by comparing the effectiveness of students using traditional text reading versus story map reading, demonstrating that using story maps helps better understand the content and structure of texts and extract essential information. Caquard and Cartwright (2014) outlined various relationships between maps and narratives, including the use of maps to represent the spatiotemporal structure of stories and their relationship to locations, emphasizing the potential of maps as narratives and the importance of linking maps with the entire mapping process through storytelling. Wright et al. (2014) suggested that storytelling through maps which tell specific and engaging stories is becoming a new form of media, exploring how "intelligent web maps" combined with text, multimedia content and intuitive user experiences can realize the enormous potential of story maps to synthesize data and express interpretive information. Kwon et al. (2023) explored how emotions evolve with topics and geographical locations and topics on Twitter with the aim to capture the dynamic nature of emotions through time series analysis and geographic visualization.

Timeline and storyline are both essential elements used in event evolution analysis to describe the sequence of events, which is another narrative visualization method relevant to this study (Brehmer et al., 2017; Fulda et al., 2016; Nguyen et al., 2016). Havre et al. (2002) proposed the ThemeRiver visualization approach to describe the evolution of themes over time in a large corpus of documents. Specifically, they used a river metaphor to convey key concepts, with the selected theme content and theme intensity represented by the river's direction, composition, changing width, and the horizontal distance between two points on the river defining the time interval. The ThemeRiver visualization ultimately shows the evolution of themes over time. Sun et al. (2014) designed a time-based visualization tool, EvoRiver, to display the dynamic changes in topics from competition to collaboration during the 2012 US presidential election period in a Twitter dataset. This visualization allows users to explore interactive experiences related to cooperation and competition and detect patterns of dynamic evolution and their leading causes. Feng et al. (2023) presented the evolution of themes in Chinese social media data during the COVID-19 pandemic using an event evolution graph created by TopicBubbler, combined with background information to present a coherent and readable storyline, achieving the research goal of transforming fragmented social media information into a complete story.

Based on the characteristics of the collected Twitter data, this study utilizes geographic location combined with information such as topics and sentiments to create a story map. In addition, by exploring the evolution of hierarchical structure themes over time, fragmented information is presented as a coherent and readable narrative.

## 3. Data and Methodology

This section presents the research data and methods, discussing the establishment of the dataset, construction of hierarchical topic structures, presentation of the story map, and event evolution analysis. Figure 1 provides the entire framework of the study from twitter data to narrative visualization.

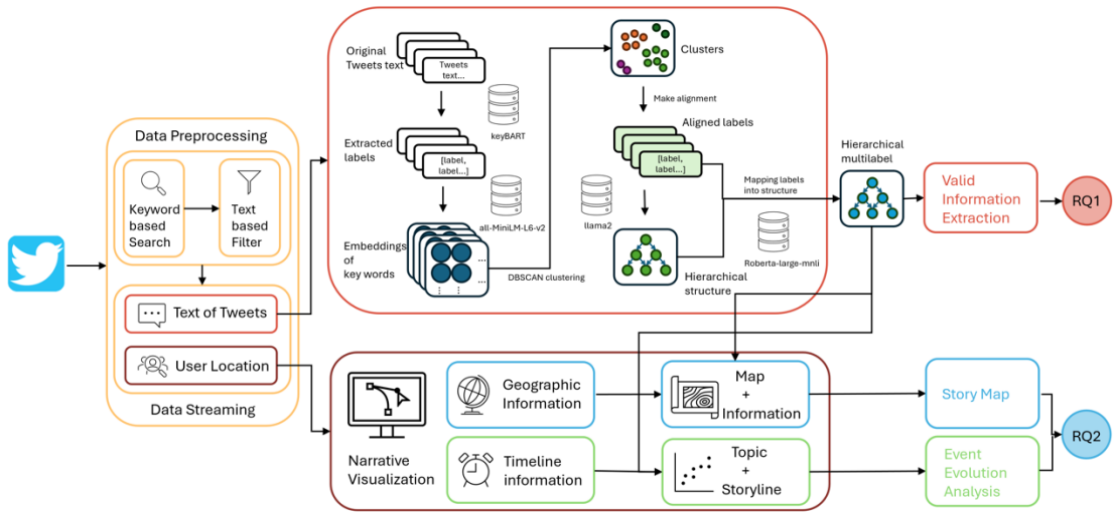

Figure 1. The Research Framework from Twitter Data to Narrative Visualization

### 3.1 Dataset Construction

To collect tweets related to AI ethics, we used topic tags to identify discourse communities revolving around specific topics (Chen et al., 2023; Jost et al., 2018). We selected seven synonymous expressions (#AI ethics, #Artificial Intelligence ethics, #Ethics of AI, #ai ethics, #Ethics In AI, #Ethical AI) to focus on AI ethics in general rather than a single ethical theme, and collected tweets from 1 January 2015 to 31 December 2022, using Python and the Twitter API (N=539,743). Only English-language tweets were considered, including text, user location, posting time, user occupation, user verification, and hashtag. We geocoded the entire tweet dataset, converted textual location descriptions into geographical coordinates, and constructed a structured geographic spatial dataset.

### 3.2 Methodology

To address the two research questions, we conducted tasks of hierarchical topic extraction, sentiment analysis, and narrative visualization of the data.

#### 3.2.1 Hierarchical Topic Extraction

Traditional topic extraction methods often require extensive data preprocessing and do not allow for creating hierarchically structured topics. We adopt an approach based on the combination of neural networks and large language models to build hierarchically structured topics.

First, the raw text data is fed into the KeyBART model (Kulkarni et al., 2022), and highly summarized vital phrases are generated by setting the num_beams parameter to 10. The value of num_beams determines the number of candidate sequences considered at each step. Typically, the value of num_beams is set between 3 and 10. Generally, a larger num_beams can improve the quality of the generated text. The words within each phrase were then lexically reduced using NLTK's grammar reduction library to align tags with similar grammatical morphology. Next, the reduced labels were converted to vectors using the all-MiniLM-L6-v2 model (Reimers and Gurevych, 2019), and density-based spatial clustering was applied to the application using a noise clustering approach to generate initial clusters (Schubert et al., 2017). Next, the aligned labels are fed into the LLAMA2 model to generate a hierarchical labeling structure (Touvron et al., 2023), and the labels are mapped to the bottom layer in the structure using the RoBERTa-large-mnli model (Liu et al., 2019). Finally, 64 labels at the bottom layer detected in the given dataset are extracted as fine-grained topics. With the hierarchical structure, the percentage of the number of original tweets corresponding to each topic can be calculated. The method realizes the conversion from raw text data to a hierarchical tag structure. In the process, multiple models and algorithms are utilized to align semantically similar tags, thus achieving efficient processing and analysis of the dataset.

### 3.2.2 Sentiment Analysis

To better illustrate the story map related to AI ethics, we employ the TweetNLP integrated platform for sentiment analysis of Twitter data (Camacho-collados et al., 2022). The core of TweetNLP is based on Transformer language models, which no longer rely on generic models or train language models from scratch but continue the training from RoBERTa and XLM-R checkpoints on Twitter-specific corpora (Conneau et al., 2020; Liu et al., 2019), providing more reliable analysis results (Devlin et al., 2019; Nguyen et al., 2020).

The sentiment analysis of TweetNLP aims to predict the sentiment of a tweet, including three labels: positive, neutral, or negative. Through TweetNLP, we predict the sentiment class of each tweet and provide a score to explain the confidence of the prediction. Moreover, tweets categorized with sentiment are further used to extract keywords of positive or negative sentiments, which helps to analyze the reasons for different sentiments.

### 3.2.3 Narrative Visualization with Data

**Story mapping**. To extract richer and more meaningful information from Twitter data, we visualize the data on maps to reveal hidden spatial narratives. Story maps intertwine geographic locations with relevant information to narrate spatial distribution stories. This paper primarily maps tweets related to AI ethics, including their topics and sentiments, onto corresponding spatial locations, supplemented with relevant text and image information to present a story map of AI ethics.

**Event evolution view**. The data processing of Twitter data typically yields fragmented information, making it challenging to form a coherent and understandable narrative. Therefore, we construct an event evolution view to integrate scattered information into a comprehensive story that allows the investigation of the development and significance of the AI ethics discourse. Since fragmented social media information may overlook deep semantic understanding when analyzed merely by computing relevance or highest frequency, we employ semantic similarity calculation to analyze the evolution of AI ethics discourse. We select specific keywords as starting words, such as the keywords appearing most frequently in January 2015, and calculate the top five keywords with the highest semantic similarity to the starting word annually as its evolutionary words. This paper utilizes the all-MiniLM-L6-V2 model to convert text labels into vectors (Reimers and Gurevych, 2019), and then use vector representations to compute the similarity between two text fragments, with the cosine similarity formula as follows:

$$Cosine\ Similarity(A, B) = \frac{A \cdot B}{||A|| ||B||} \qquad (1)$$

where A and B are the vector representations of the two text labels, · denotes the dot product of the vectors, and ||A|| and ||B|| are the norms (i.e., lengths) of the vectors A and B, respectively.

## 4. Results

The findings of this paper consist of an overview of the hierarchical topic structure, a story map with geographic locations, and an event evolution view integrating fragmented information.

### 4.1 Overview of Hierarchical Topic Structure

To address our first research question (RQ1), about how AI ethics discourse is framed within Twitter discourse overall, we extracted all relevant tweets and clustered them into a hierarchical topic structure, as shown in Figure 2. This structure consists of three layers. The first layer comprises seven main topics: Legal & Ethical, Society & Culture, Technology, Science & Research, Health & Safety, Education & Learning, and Business & Economics. Underneath each of the seven top-level topics are subcategories at the second and the third layer (bottom layer), which presents 64 fine-grained topics covering mainstream public discourse and issues of small but critical scope.

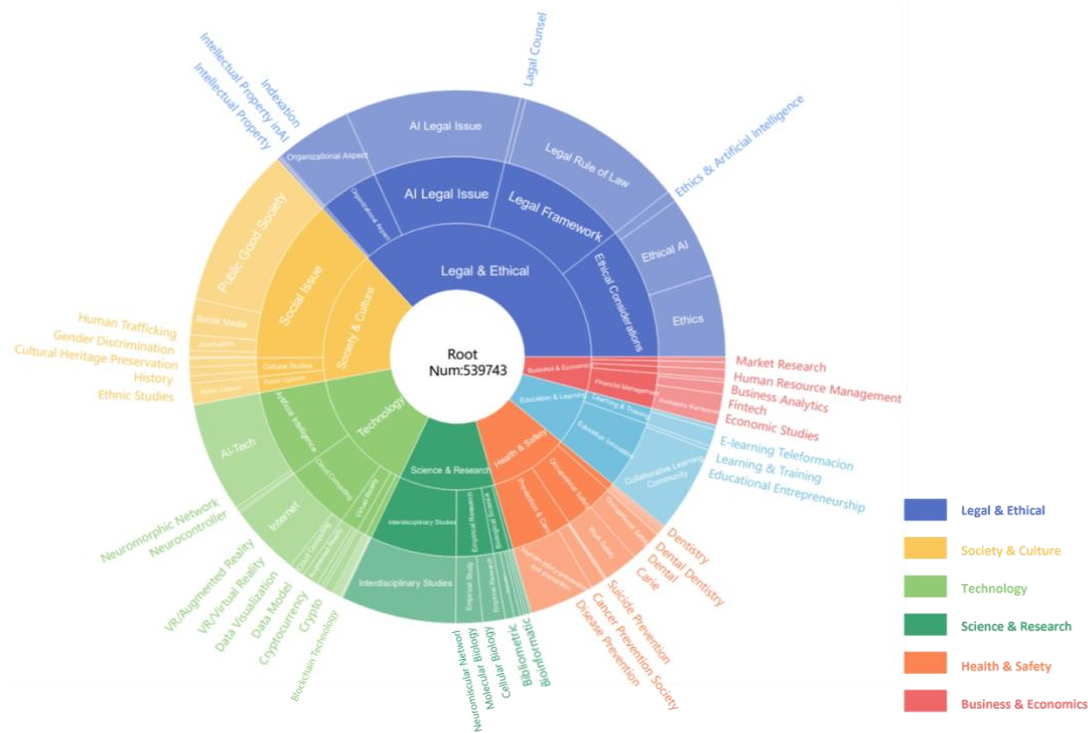

Figure 2. A Hierarchical Thematic Structure of AI Ethics Twitter Discourse.

In addition to the fine-grained hierarchical structure analysis, the temporal trend of topic discussions over time also conveys rich meaning. we chose stream chart and bar chart to illustrate the changes of the seven main topics related to AI ethics from 1 January 2015 to 31 December 2022 as Figure 3A shows. The overall volume of discussions on AI ethics was relatively low during 2015-2016, gradually increasing from 2017 and peaking in early 2020. There was a slight decline in 2020, possibly influenced by the pandemic outbreak, which may have diverted substantial discourse resources. After 2020, the discourse on AI ethics remained relatively stable. Figure 3B shows that among the seven topics, a significant portion of the discussions revolve around "Legal & Ethical," surpassing 90% of the total tweet volume. The remaining six topics have relatively similar amount of discussion, with "Business & Economics" being the least discussed. This skewed distribution, where a few categories (also called heads) contain a large number of samples, while most categories (also called tails) have very few samples, conforms to a long-tail distribution (Anderson, 2012). The long-tail distribution of topics related to AI ethics reveals that although public concern in this field is focused on Legal & Ethical discussions, niche topics such as Education & Learning and Business & Economics are also significant and should not be overlooked (Agarwal et al., 2012; Mustafaraj et al., 2011).

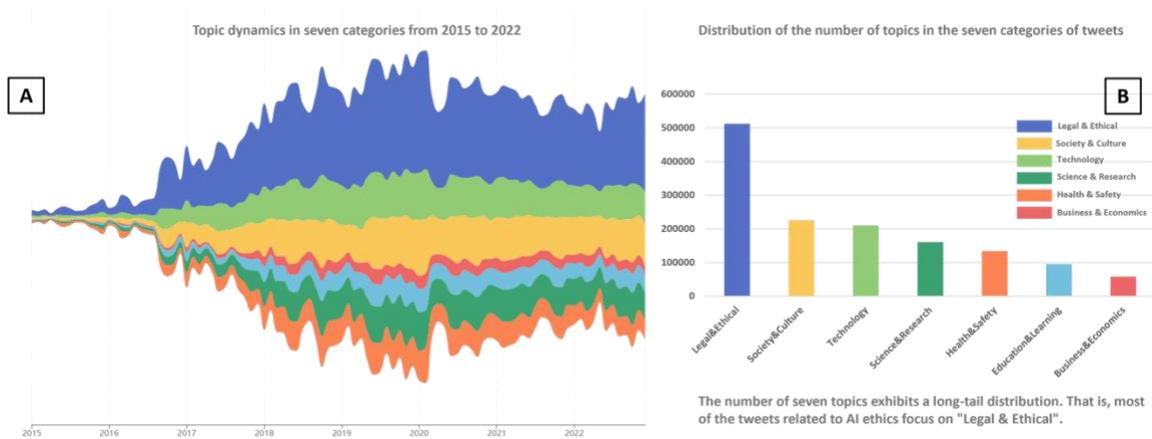

Figure 3. Temporal Trend and Distribution of AI Ethics Topics on Twitter.

### 4.2 Narrative visualization of AI ethics discourse in twitter

To answer RQ2, this study presents a coherent and readable narrative visualization from two parts, including story map and event evolution diagram.

### 4.2.1 Story Map: The World and The United States as Examples

We constructed a global story map and a more detailed story map of the United States, which has the highest tweet volume, as an example. Among them, the global story map combines mainstream AI ethics discourse with geospatial information, presenting the distribution of seven topics worldwide. The American Story Map integrates mainstream AI ethics discourse, sentiment information, and spatial locations, built upon the changes in the number of tweets related to AI ethics in the United States from 2015 to 2022.

Figure 4 illustrates the distribution of AI ethics-related tweets worldwide. Most AI ethics discourse is concentrated in the United States and Europe. Some countries, such as China and Cuba, have limited use of Twitter, so their distribution data may not provide accurate references. The surrounding maps in Figure 4 display the distribution of the seven topics worldwide, they are generally similar but with some subtle differences. For instance, discussions on "Health & Safety" in African countries are more prevalent compared to "Technology" and "Business & Economics," while India has more discussions on "Technology" and "Education & Learning" compared to other topics.

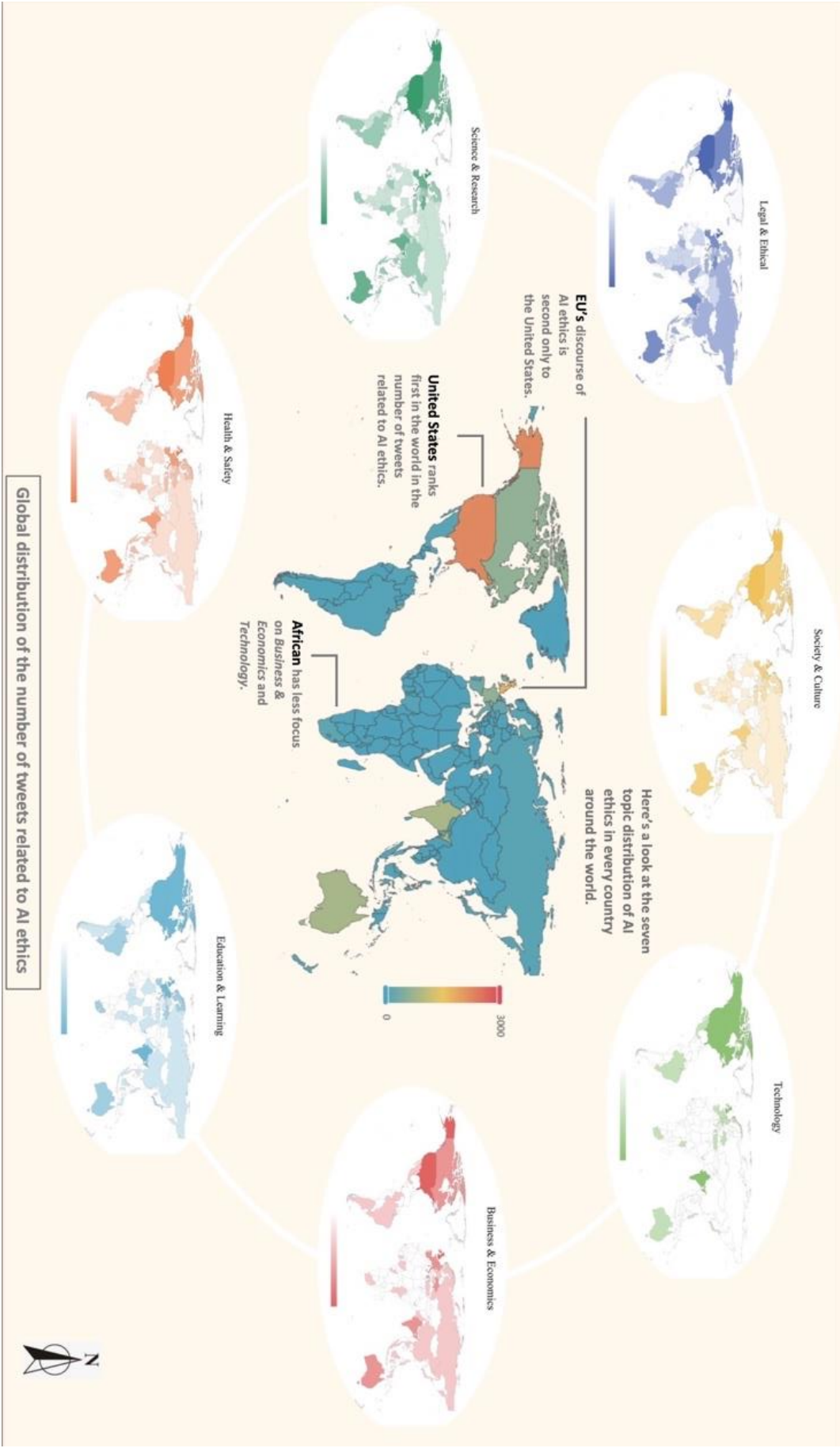

Figure 4. Global Distribution of AI Ethics Twitter Data and Distribution of Seven Topics Worldwide.

To analyze the information conveyed by the story map in Twitter data further, we present an example using the United States. Figure 5A displays the distribution of AI ethics-related tweets in the United States. We observe that the discussion of AI ethics topics is most concentrated in California, New York, and Massachusetts. This concentration might be attributed to the large population size, numerous high-tech companies, and the abundance of universities in these three states. Considering the "long-tail distribution" characteristic of the seven topics related to AI ethics, although the tail-end topics constitute a relatively small proportion, they still reveal significant information. After excluding the most discussed topic "Legal & Ethical," we illustrate the distribution of the remaining topics in Figure 5B. More than 50% of states focus on "Science & Culture," with "Technology" as the next prominent topic. Interestingly, New Mexico is most interested in the "Health & Safety" topic. This may be due to the diverse social and cultural backgrounds of the different federal states. For example, California and Washington State are home to numerous large tech companies. Still, California's industries include globally renowned tourism and film industries, while Washington State is known for aerospace and agriculture, resulting in differing AI ethics discourse between the two states.

Figures 5C and 5D present the distribution of positive and negative sentiments related to AI ethics discourse in various states. Interestingly, the top five states with the strongest positive sentiment are the same as the top five states with the strongest negative sentiment: California, New York, Massachusetts, Washington, and Texas. This result reflects the consistent intensity of public sentiment; regions expressing positive sentiments do not necessarily have reduced negative sentiments. Furthermore, we explored the content discussed behind these positive and negative sentiments and displayed them using word clouds. Figures 5E and 5F show the word clouds corresponding to positive sentiment and negative sentiment respectively. This indicates that people discuss similar topics with different sentiments, focusing on data, humans, and artificial intelligence. However, those expressing positive sentiments are more likely to see the positive impacts of data and AI on humanity, while those expressing negative sentiments demonstrate more ethical concerns and are also more concerned about the potential problems with data. Figure 5G provides statistical information on the trend of AI ethics discourse in the United States over time, serving as background information for the story map.

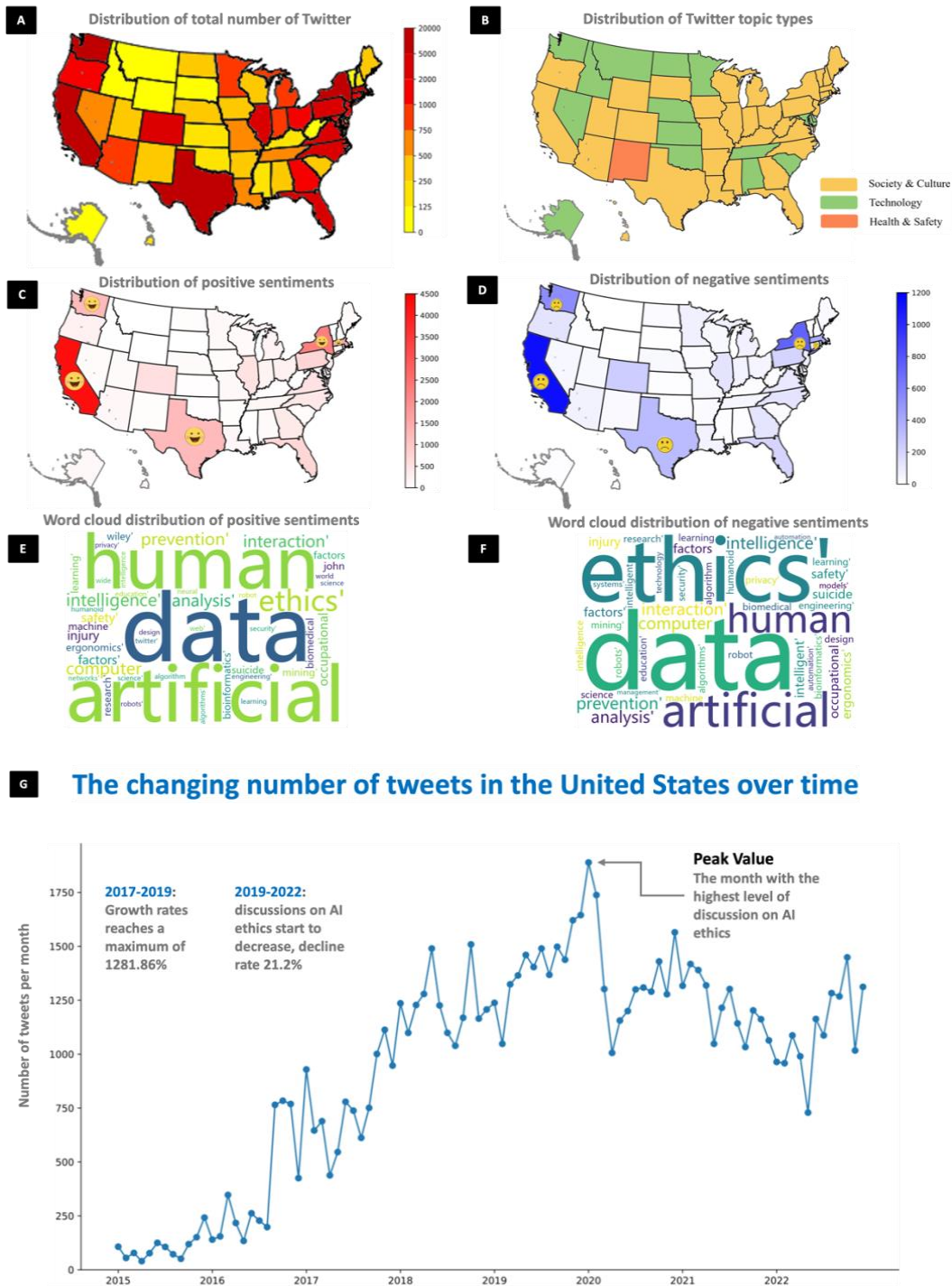

Figure 5. Story Maps of AI Ethics Discourse in the United States.

4.2.2 Event Evolution View

We integrated fragmented AI ethics discourse information into readable and coherent narratives. Figure 6A depicts the evolution of AI ethics discourse worldwide over time, with some critical events related to AI as background information. Figures 6B and 6C illustrate the evolution of AI ethics discourse from 2015 to 2022. The horizontal axis represents the timeline, while the vertical axis indicates the magnitude of each topic. Since this event evolution diagram aims to display the evolution of topics, time and quantity serve as reference information for the distribution of topic bubbles. We first extracted the main topics from the discourse in January 2015, identifying "Internet" as the most discussed one in that month. Then, by calculating semantic similarity, we selected five topics semantically most related to "Internet" in 2015, such as AI-tech, Crypto, etc. The topic with the latest timestamp in the previous year then evolved into the following year's five topics, and so on.

Through the consolidation of fragmented information related to AI ethics from 2015 to 2022, combined with background information, we present an event evolution view. In 2015, the formal launch of Ethereum and incidents such as hacking of Bitcoin exchanges sparked discussions on "Crypto" and "Cryptocurrency" related to AI technology. In 2016, following the theft of Ether from The DAO, an Ethereum-based intelligent contract organization, attention towards financial technology continued to rise. Blockchain technology gradually began to be used in the financial sector to ensure security and compliance. In 2017, the continued development of AI technology drew multidisciplinary attention. Single-discipline advancements were no longer sufficient to address complex real-world issues, leading to an emphasis on interdisciplinary research. In 2018, public discussions on ethics reached unprecedented levels. In May of the same year, the European Union formally implemented the General Data Protection Regulation (GDPR), setting a benchmark for AI ethics regulations. In addition to "Ethics," the public showed increased interest in cultural and economic-related topics. In 2019, the trade war triggered fluctuations in the global economic market. Bitcoin plummeted, leading to widespread closures of mining operations. The strengthening of the US dollar and continuous interest rate hikes by the Federal Reserve caused emerging market currencies to collapse. Topics such as "Economic Studies," "Business Analytics," and "Market Research" became focal points of discussion. In 2020, the outbreak of COVID-19 consumed a significant portion of social media resources, resulting in fewer discussions related to AI ethics. As the pandemic spread, many countries implemented surveillance and tracking measures to control its spread, leading to ethical debates on privacy and surveillance. The pandemic-induced home isolation and remote work shifted learning and work patterns, making topics like "Ethics & Artificial Intelligence," "Collaborative Learning Community," and "E-Learning Teleformacion" new focal points of discussion. In 2021, the global gaming platform Roblox became the first metaverse concept stock listed on the New York Stock Exchange, sparking discussions on "Augmented Reality" and "Virtual Reality." In April of the same year, the European Union proposed the Artificial Intelligence Act (EU AI Act), the world's first comprehensive legislative attempt to address the phenomenon and risks of artificial intelligence. The establishment of this AI regulatory framework led to an increase in discussions on "Ethical AI." In 2022, DeepMind successfully predicted the structures of approximately 200 million proteins from 1 million species using AlphaFold, covering almost all known proteins on Earth, ushering humanity into a new era of digital biology. In April, the international academic journal Science revealed the mystery of human genes, announcing the completion of the first complete map of the human genome. Topics such as "Neuromuscular Network," "Molecular Biology," "Cellular Biology," and "Ergonomic Design" became focal points of discussion.

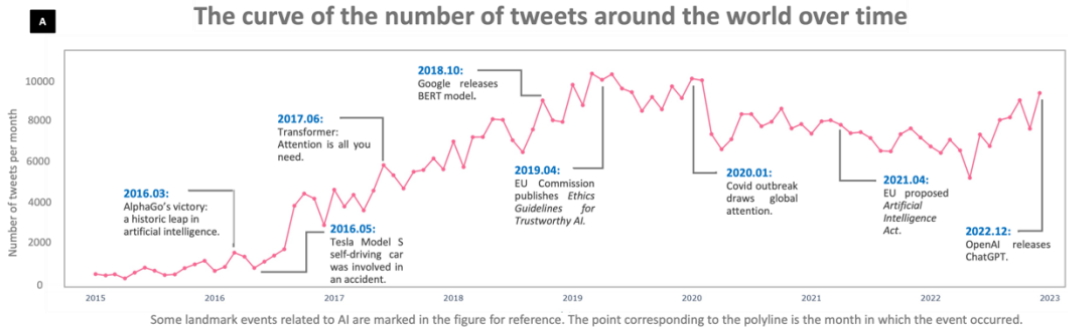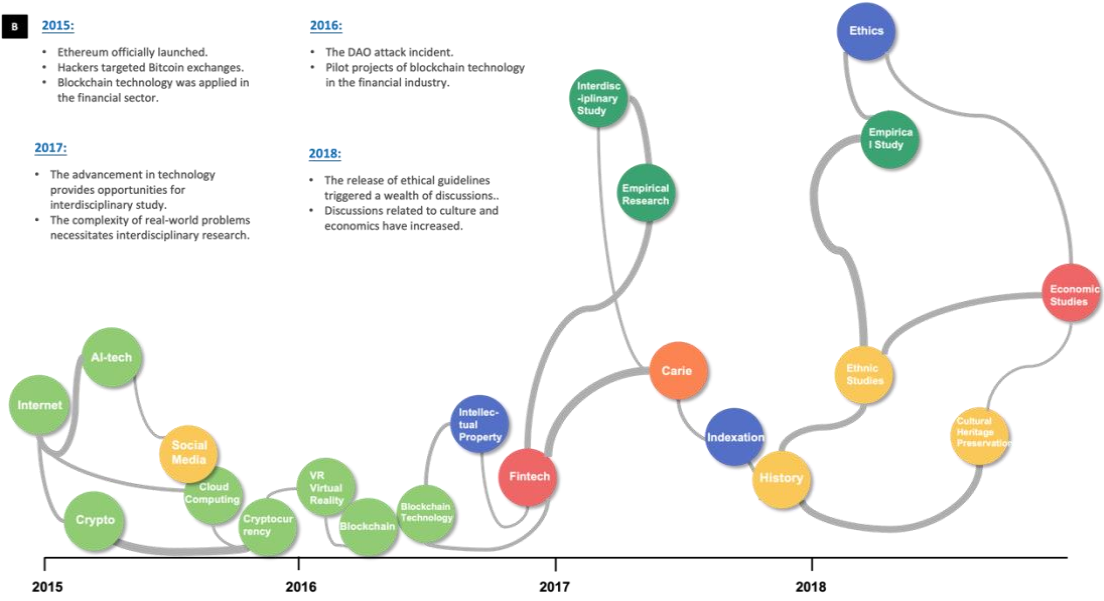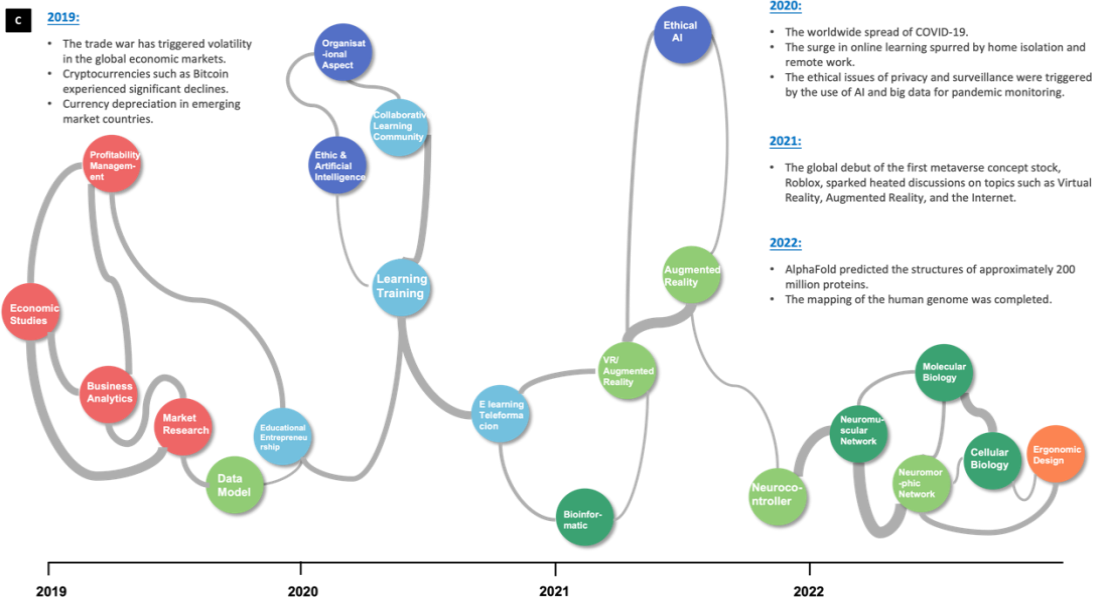

Figure 6. Event Evolution View of AI Ethics Discourse on Twitter.

## 5. Discussion

5.1 The Hierarchically Structured Topic Framework of AI Ethics Discourse in Twitter

AI ethics frameworks have been continuously proposed, either focusing on a single domain (Holmes et al., 2022; Vakkuri et al., 2019), based on policymakers' perspectives (Floridi and Cowls, 2022), or constructed from the Wikipedia platform to build the topic structure of AI ethics (Wei et al., 2024). However, hierarchically structured topic frameworks are rarely constructed to study AI ethics discourse. On the one hand, it is challenging to collect a large amount of data from the public perspective. On the other hand, information on social media platforms is often flooded with interferences and noise, making it difficult to form valuable, structured information. This paper combines neural networks with large language models to extract tweet topics related to AI ethics, effectively filtering out noise interference and constructing AI ethics discourse framework on Twitter. The seven topics at top level encompass all categories of AI ethics discourse. The 64 labels at bottom level reveal the fine-grained topic distribution. Such a hierarchical topic structure provides a new perspective on AI ethics discourse based on the public perspective.

## 5.2 Long Tail Distribution in Social Media

Our research also values small but critical voices in AI ethics discourse on Twitter. Social media data is typically massive and multidimensional, and the dissemination of information is easily guided by primary talkers so that much of the discussion revolves around the dominant voices. Our results demonstrate that the distribution of AI ethics discourse on Twitter follows a "long-tail distribution," with a significant focus on "Legal & Ethical," while the discussion amount of the remaining six topics is relatively small. Chris Anderson (2006) analyzed the business and economic models of Amazon, Netflix, and other websites. He found that small but diverse products or services that were underappreciated had cumulatively outperformed mainstream products due to their large volume. Similarly, in the "long-tail distribution" of AI ethics discourse, apart from "Legal & Ethical," the remaining six topics, although having lower discussion volumes, still reveal many meaningful insights. For example, in New Mexico in the United States, the hottest discussed topic in AI ethics discourse is "Health & safety", which may be attributed to its local economy, and people are more concerned about living necessities.

## 5.3 The Event Evolution of the AI Ethics Discourse

Our research uses narrative visualization to integrate fragmented information into an event evolution view, thus allows to explore the hidden events in social media data. We present a complete story by combining significant societal events with background information. This not only helps us understand the evolution of AI ethics discourse but also provides new insights into how major societal events influence AI ethics discourse. For example, the emergence of COVID-19 in 2020 led to a decrease in AI ethics discourse. Meanwhile, home isolation and remote work prompted the emergence of new forms of learning, such as e-learning and collaborative learning.

# 6. Limitation

Although our results have effectively answered the two research questions, they still have some limitations with regard to data and methods. First, our approach of crawling data using keywords related to AI ethics may not fully capture all relevant tweets. For instance, content without AI ethics-related hashtags added by the tweet publishers would not be retrieved. Moreover, the public's use of social media platforms is unbalanced. For example, it may be more difficult for economically underdeveloped regions to use social media platforms, resulting in limited accessible data. Second, there are limitations in selecting topics in the AI ethics event evolution view. The method reported in this paper is just one approach that helps us select evolving topics based on semantic similarity. Other methods, such as frequency-based or intensity-based selection, could also be considered. Moreover, this paper focuses on the extension and evolution of AI ethics discourse, so we excluded repeatedly occurring old topics and ultimately selected only the five new topics with the highest semantic relevance. Third, the representation of the AI ethics event evolution view is limited. The evolution of AI ethics events should ideally include information such as the time and volume of discussion for each topic. However, our results involve exploring topics at a granular level, leading to significant differences in discussion volume for each topic. Additionally, restrictions in visualization effects and methods make it difficult to accurately represent the temporal distribution of topics within a constrained space. What we have created is only an illustrative view of the evolution of AI ethics discourse by combining time and discussion volume for each topic.

## 7. Conclusion and Outlook

By exploring AI ethics discourse on Twitter, we constructed fine-grained hierarchical structure topics, and integrated fragmented information into narratives containing story maps and an event evolution view. Our research aims to facilitate active public oversight of AI technologies for their fair and sustainable development. The contributions of the study are twofold. First, we applied hierarchically structured topic framework to construct social media discourse on AI ethics, which differs from traditional social media research focused only on trendy topics, and achieves the goal of fine-grained exploration of social media data. Second, we integrated fragmented social media information into a readable and comprehensive narrative. This helps the public to better understand and participate in the construction of the AI ethics discourse.

The exploration of AI ethics and social media data is just at the beginning. The essence of AI technology is to serve humanity, and the public ethical concerns and discussions sparked by AI are worthy of attention. For example, future research could combine news and Twitter data to analyze AI ethics issues. Investigating public discussions and feedback on Twitter regarding a specific AI ethics news case can help policymakers formulate more specific guidance to address and mitigate ethical concerns from the public perspective.